**Risk Assessment of COVID Infection by Respiratory Droplets from Cough for Various Ventilation Scenarios Inside an Elevator: An OpenFOAM based Computational Fluid Dynamics Analysis**

Riddhideep Biswas[1], Anish Pal[1], Ritam Pal[2], Sourav Sarkar[1,*], Achintya Mukhopadhyay[1]

[1]Department of Mechanical Engineering, Jadavpur University, Kolkata-700032, India

[2]Department of Mechanical Engineering, The Pennsylvania State University, University Park, Pennsylvania 16802, U.S.A.

*Corresponding author: souravsarkar.mech@jadavpuruniversity.in

**Abstract**

Respiratory droplets which may contain the disease spreading virus, exhaled during speaking, coughing or sneezing are one of significant causes for the spread of the ongoing Covid-19 pandemic. The droplet dispersion depends on the surrounding air velocity, ambient temperature and relative humidity. In a confined space like an elevator, the risk of transmission becomes higher when there is an infected person inside the elevator with other individuals. In this work, a numerical investigation is carried out in a three-dimensional domain resembling an elevator using OpenFoam. Three different modes of air ventilation i.e quiescent, axial exhaust draft and exhaust fan have been considered to investigate the effect of ventilation on droplet transmission for two different climatic conditions (30℃, 50% relative humidity and 10℃ , 90% relative humidity). The risk assessment is quantified using a risk factor based on the time-averaged droplet count present near the passenger's hand to head region (risky height zone). The risk factor drops from 40% in quiescent scenario to 0% in exhaust fan ventilation condition in a hot dry environment. In general, the cold humid condition is safer than the hot dry condition as the droplets settle down quickly below the risky height zone owing to their larger masses maintained by negligible evaporation. However, the exhaust fan renders the domain in hot dry ambience completely safe (Risk factor, 0%) in 5.5s whereas it takes 7.48s for cold humid ambience.

**I. Introduction**

The ongoing pandemic of Covid-19 has caused an immense loss of lives as well as shattered the global economy, the entire world still suffering from its excruciating shackles. The transmission of Sars-Cov-2 has been responsible for this pandemic and it has been established that the virus spreads through respiratory droplets[1-3]. Respiratory droplets are exhaled from the mouth or nose during speaking, coughing or sneezing. The conditions of speaking, coughing or sneezing can be differentiated with the help of droplet size and velocity spectra and exhalation velocity of jet from the mouth[4,5]. The respiratory droplets expelled from the mouth by coughing or sneezing, are polydisperse in nature. Kwon et al. performed experiments involving particle image velocimetry to study the initial velocity distributions from coughing and sneezing[6]. Johnson et al. performed experiments with Aerodynamic Particle Sizer in order to obtain the aerosol size distribution during coughing[7]. Li et al. computationally modelled the evaporation of cough droplets employing the multi-component Eulerian-Lagrangian approach in inhomogeneous humidity condition[8]. It is expected that the droplets will follow a projectile motion and eventually settle on the ground. However, during coughing, a turbulence is generated by the air puff which takes the responsibility of carrying the droplets an appreciable amount of distance. While travelling the distance, the droplet undergoes evaporation which reduces the size of the droplets thereby helping them to cover more distance. Redrow et al. modelled the evaporation and dispersion of cough jet droplets considering sputum for the first time and showed that the human cough includes a turbulent air puff[9]. Wei et al. showed that the droplet spread gets enhanced by the presence of turbulent air jet[10]. Pal et al. investigated the influence of ambient conditions on droplet transport in indoor environments. The droplet trajectories were calculated considering a turbulent round jet model[11]. Chong et al. quantified the lifetime of respiratory droplets under different ambient humidity conditions using direct numerical solution (DNS). They have reported that the droplet lifetime can extend even up to 150 times at very high relative humidity (90%)[12]. The droplet size can increase initially in cold environments due to the supersaturation of vapor turbulent puff [13]. Wang et al. performed flow visualization and particle image velocimetry to understand the motion of the droplets and utilised those results to establish a physical model for investigating the trajectories of the droplets expelled by cough[14]. Pendar et al. conducted a wide range study of the velocity distribution, expelled angle and the size of the droplets released from the mouth in order to determine the correct social distance guidelines for different conditions[15]. Dbouk and Drikakis have conducted a computational study to investigate the transport, dispersion and evaporation of respiratory droplets produced from a human cough[16]. If a person is present in an open area like a school ground or market place, the risk of the disease transmission depends on wind velocity due to the available space between two persons. Feng et al. carried out a numerical study to investigate the effects of wind and relative humidity on the transportation of respiratory droplets in an outdoor environment for different wind velocity ranges from 0-16 km/h[17]. They have found that the droplet cloud can travel up to 8 m in the presence of wind. Li et al. performed numerical study on dispersion of cough droplets with non-volatile components in a tropical outdoor environment and the dispersion was influenced by the relative humidity and wind speed[18].

In an open space, it has been observed from the above literatures that the wind speed affects the spread of the droplets but the situation becomes alarming in a confined space where the environment is enclosed and if a person coughs, the droplets will remain inside the space for an appreciable amount of time. In recent times several studies have been performed on aerosol and droplet transmission in different confined spaces like classroom[19], aircraft cabin [20,21], restaurants [22,23], bus[24], clinic[25] etc. Liu et al. performed a laboratory study to investigate the expiratory airflow and particle dispersion in a stratified indoor environment[26]. Cheng et al. investigated the trajectories of large respiratory droplets in indoor environments under different relative humidity[27]. They reported that droplets produced from coughing can travel distances of 1.09 m and that of sneezing can cover a distance of 2.76 m. Yang et al. carried out a computational study to capture the dispersion of pathogen-laden respiratory droplets in an enclosed environment like crowded bus[24]. Foster et al. studied infection probability in the classroom scenario with masked habitants and different ventilation conditions[19]. Yan et al. employed the Lagrangian-based Wells-Riley approach to assess the risk of airborne disease infection in an airliner cabin[21]. Yan et al. numerically investigated the thermal effects of the human body on the evaporation and dispersion of cough droplets in an enclosed environment[28]. Sen performed a numerical study to investigate the evaporation and transmission of cough droplets in a confined space like elevator considering different scenarios such as varying the air ventilation systems, number of persons inside the elevator, direction of ejection, relative humidity and temperature[29]. Dbouk et al. showed how the modifications in ventilation systems in confined spaces can influence the transmission of airborne virus[30]. Agrawal et al. have worked on the reduction of the risk associated with cough clouds in a closed space by modifying the ventilation systems[31]. It was found that the infected air volume is 23 times the ejected volume of air by cough.

All these works have paid attention to the ventilation systems but none of them have studied the risk to which another person will be exposed when he or she tries to board the elevator after the door opens. The respiratory droplets are laden with salt and proteins which contain the virions. Moreover, in most of these works, pure water droplets have been considered which does not resemble a real scenario where the droplet is pathogen laden. The droplets contain non-volatile salts in some specific proportions and these salts contain the pathogen. A salt laden droplet will have thermophysical properties different from that of pure water droplet, and the thermophysical properties will constantly change with evaporation, unlike pure water. This difference and constant change in thermophysical properties will cause a difference in the mass transfer number and hence the evaporation rate, ultimately manifesting itself in a difference in the overall droplet dispersion and trajectory than that of pure water. So, in order to mitigate the above stated problem and to make the simulations and their corresponding results more realistic, our work has implemented this salt model of droplet along with droplet evaporation. In addition, a systematic comparison of the risk of infection from different designs of ventilation along with the safety measures to be adopted in each of the designs has been carried out in this work.

In confined spaces, the ventilation plays a significant role which has been highlighted in the existing literature. When the droplets are inside the enclosed space, they will undergo evaporation which will transform them into aerosol and they can remain suspended in the air for longer periods of time. Several works have been reported recently where the aerosol route of virus transmission has been supported with enough evidences[16,32-35]. The aerosol size is generally considered less than 10 micron which helps them to remain suspended in the air for longer periods of time. Furthermore from the recent works, it can be concluded that the aerosol constitutes the virions which are responsible for the disease transmission[36]. So, when a person coughs or sneezes in a confined space like an elevator, the aerosol will remain in the space until it is forced out or it gets any opening to discharge outside. This implies that when an elevator door will open for the passenger who is waiting outside, the passenger will be exposed to a high-risk situation in the elevator if the former passenger inside the lift has just coughed before opening of the door. Face masks may work as a protective gear up to a certain extent but there are limitations. Akhtar et al. performed droplet flow visualization experiments to test the effectiveness of five different masks and it was found that except for the N-95 masks, all the other masks showed some amount of droplet leakage[37]. Dbouk et al. have also worked on the effectiveness of masks to reduce the droplet transmission[33]. The risk assessment of such situations is necessary as it will help to understand the safe time interval between stopping of the elevator and boarding it. Moreover, the assessment can also give information regarding the safe distance which must be maintained from the infected passenger inside the elevator.

In this work, a three-dimensional simulation has been performed to study droplet transmission inside an elevator. An infected passenger is present inside that elevator without a face-mask and the passenger is coughing. Different ventilation conditions of the ambience have been investigated considering both quiescent environment and forced circulation of air in the domain. The dispersion and evaporation of the droplets (including non-volatile components) has been investigated in each of the cases. The situation is such that the person has boarded the elevator and before alighting from the elevator, the person coughs. The beleaguered person waiting outside to board the elevator will be oblivious to the threat that looms inside the elevator. Our paper focuses on the risk assessment of this situation for various ventilation conditions along with suggestions of safety measures.

## II. Problem Formulation

### A. Geometry

The computational domain consists of an elevator having a capacity of 5 persons which is portrayed in Fig. 1(a) , describing its details. The dimension of the elevator is 1.2 X 1.2 X 2m, which represents a characteristic size of elevators present in housing complexes or small enterprises. Investigation of an elevator of relatively smaller size is incumbent as the risk associated with a smaller confined space is higher. At the top of the elevator, a circular mounting of 0.6 m diameter is provided. This top circular mounting is subjected to conditions specific to each particular scenario, as illustrated further in the Initial and Boundary Conditions section as well as in Table 1. This top mounting is a very important part as by altering the boundary conditions on this mounting, the various ventilation situations inside the elevator are realized. In the present study, boundary conditions representing a axial exhaust jet or an exhaust fan at the top opening have been presented. The ventilation slots, 3% of the platform area, have been provided at the lower portion of the elevator side walls (complying with the European EN-81-1 code[38]). Instead of incorporating a human manikin, the features of the passenger namely, the head, face, mouth and remaining body parts have been represented with rectangular boxes, whose dimensions conform to that of an actual human being in an upright standing position, to reduce the complexity and the computational cost and time. Also, since our area of focus in the domain is far away from the passenger, implementation of a geometry that corresponds to an actual human being will have no effect on our desired zone of interest. The passenger height is taken as 1.75m (the mouth height being 1.56 m)[39]. Figure 1(b) shows the isometric view of the passenger in the domain, which also shows the position of the passenger relative to the elevator walls. The mouth of the passenger, from which the cough droplets are injected, has been modelled as a rectangular-shaped aperture having a width of 40 mm and an aspect ratio of 8[39], as shown in Fig. 1(c) . Three different scenarios have been studied to comprehend the effect of air flow in the surrounding environment on transmission and evaporation of droplets, injected into the domain by coughing of the passenger. Droplets ejected in the domain due to the coughing of the passenger would traverse the domain, their trajectory being contingent upon the preponderant velocity field. Certain number of droplets could turn into aerosols based on the prevalent temperature and humidity. A number of droplets might escape, a few of them may stick on the various surfaces within the elevator, while the remaining ones will remain suspended in the domain for extended periods – the last ones being of major concern to us.

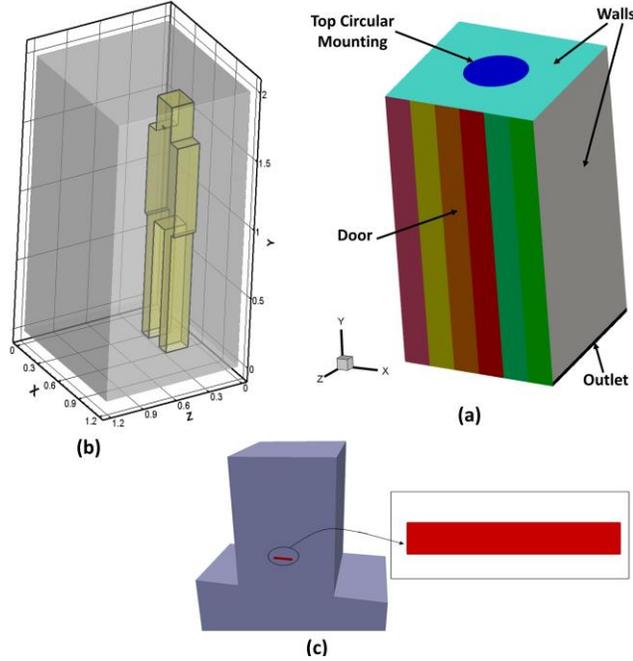

**Fig. 1 : (a) The whole computational domain.
(b) Isometric view of the passenger in the domain.
(c) Mouth of the passenger modelled as a rectangular aperture.**

**B. Mathematical Model**

An Eulerian-Lagrangian model has been implemented for this numerical study. Air, the carrier fluid is modelled in the Eulerian frame. For the carrier bulk multiphase fluid mixture, the continuity (equation 1)[29,40] and the compressible multiphase mixture Reynolds-averaged Navier–Stokes equations (equation 2)[29,40] in conjunction with the k – ω turbulence model in the shear-stress-transport formulation(SST) (equations 3-15)[42] has been employed. Droplets that are injected into the domain due to coughing are treated as discrete particles and are solved in the Lagrangian frame of reference. The droplets roam around depending on the prevailing air velocity field and undergo evaporation as they traverse the domain. The droplets and moist air at the time of ejection from the mouth can be considered to be at the same temperature as that of the body temperature. Eventually, however the cloud of droplets intermingles with the surrounding air and it draws in a substantial amount of the surrounding ambient air following which the temperature of the cloud would effectually be identical to the surrounding ambient temperature [31] . But the evaporation process still continues (until the droplet becomes devoid of the volatile component), and the energy required for the phase change to happen is attained from the droplet and the surroundings (equation 22)[29]. The driving potential for the occurrence of evaporation of water is the difference of partial pressure of water vapour at the droplet surface and in the air encompassing the droplets (equation 27)[29]. The rate of evaporation depends upon the mass transfer coefficient determined from the Sherwood number (equation 28). Sherwood number, again, is dependent upon the droplet Reynolds number (equation 29)[40] based on the Ranz-Marshall correlation (equation 28)[33,34]. The droplet is considered to be a mixture of salt and liquid water (99% water and 1% NaCl by wt.). As the droplets evaporate to lose the volatile liquid mass into the ambient (equation 18)[29] and their diameter decreases, the mass fraction of its components changes (equations 32, 33) and finally the droplet fully evaporates i.e. becomes fully devoid of the volatile liquid water component, forming droplet nuclei. The droplet properties used in the governing equations are a function of the properties of liquid water and salt as well as their respective mass fractions (equations 34, 35). The continuous change in the mass fractions of the components owing to the evaporation of the droplets has been taken into account (equations 31-33). Hence, to incorporate this salt model of the droplets (equations 31-36), modifications has been made in the source code of the reactingParcelFoam solver of OpenFOAM. The Ranz-Marshall model has been implemented to calculate the Nusselt number (equation 24)[43,44] and Sherwood number (equation 28)[43,44], to solve the droplet heat transfer (equation 22)[29] and mass transfer (equation 27)[29]. The droplet temperature is obtained by solving the energy conservation equation, as discussed below (equation 22)[29]. The evaporative cooling is modelled by taking into account the energy transfer from the bulk phase into Lagrangian phase (equations 15-17,23-24)[29]. The position and the velocity of the droplets are obtained by applying Newton's second law of motion on the droplets and the forces considered here are gravity, buoyancy, lift and drag (equations 21,25, 26)[29,40]. The relevant gas phase and particle phase transport equations with appropriate closure relations are given below.

**Governing Equations for the Eulerian (Gas) Phase:-**

Continuity equation

$$\frac{\partial \rho}{\partial t} + \frac{\partial}{\partial x_j}(\rho u_j) = m_v''' \tag{1}$$

Momentum equation

$$\rho \frac{\partial u_i}{\partial t} + \frac{\partial}{\partial x_j}(\rho u_i u_j) = -\frac{\partial p}{\partial x_j} + \frac{\partial \tau_{ij}}{\partial x_i} + \rho g_i \tag{2}$$

Closure equation for the momentum equation

$$\tau_{ij} = \mu_t \left(2 S_{ij} - \frac{2}{3}\frac{\partial u_k}{\partial x_k}\delta_{ij}\right) - \frac{2}{3}\rho k \delta_{ij} \tag{3}$$

Strain rate

$$S_{ij} = \frac{1}{2}\left(\frac{\partial u_i}{\partial x_j} + \frac{\partial u_j}{\partial x_i}\right) \tag{4}$$

Transport equation for turbulent kinetic energy

$$\frac{\partial(\rho k)}{\partial t} + \frac{\partial(\rho u_j k)}{\partial x_j} = P - \beta^* \rho \omega k + \frac{\partial}{\partial x_j}\left[(\mu + \sigma_k \mu_t)\frac{\partial k}{\partial x_j}\right] \tag{5}$$

Transport equation for turbulent energy dissipation

$$\frac{\partial(\rho k)}{\partial t} + \frac{\partial(\rho u_j k)}{\partial x_j} = P - \beta^* \rho \omega k + \frac{\partial}{\partial x_j}\left[(\mu + \sigma_k \mu_t)\frac{\partial k}{\partial x_j}\right] \tag{5}$$

Turbulent kinetic energy production

$$P = \tau_{ij}\frac{\partial u_i}{\partial x_j} \tag{7}$$

Eddy viscosity limiter

$$\mu_t = \frac{\rho a_1 k}{\max(a_1 \omega, \Omega F_2)} \tag{8}$$

Weighted model constants

$$\phi = F_1 \phi_1 + (1 - F_1)\phi_2 \tag{9}$$

Blending function 1

$$F_1 = \tanh(arg_1^4) \tag{10}$$

Argument for blending function 1

$$arg_1 = \min\left[\max\left(\frac{\sqrt{k}}{\beta^*\omega d}, \frac{500\nu}{d^2\omega}\right), \frac{4\rho\sigma_{\omega 2}\kappa}{CD_{k\omega}d^2}\right] \qquad (11)$$

Blending function 2

$$F_2 = \tanh(arg_2^2) \qquad (12)$$

Argument for blending function 2

$$arg_2 = \max\left(2\frac{\sqrt{k}}{\beta^*\omega d}, \frac{500\nu}{d^2\omega}\right) \qquad (13)$$

Constants

$$\left.\begin{array}{l} \sigma_{k1} = 0.85, \sigma_{w1} = 0.65, \beta_1 = 0.075 \\ \sigma_{k2} = 1.00, \sigma_{w2} = 0.856, \beta_2 = 0.0828 \\ \beta^* = 0.09, a_1 = 0.31 \end{array}\right\} \qquad (14)$$

Energy transport from Lagrangian to Eulerian phase

$$\rho\frac{\partial H}{\partial t} + \frac{\partial}{\partial x_i}\left[\rho u_i H - \frac{\partial}{\partial x_i}(k_{eff}T)\right] = Q_d \qquad (15)$$

Closure term for Energy equation

$$H = \sum x_i C_{pi} T_i \qquad (16)$$

$$K_{eff} = k_{mol} + k_t \qquad (17)$$

Species transport equation for water vapour

$$\frac{\partial(\rho f_v)}{\partial t} + \frac{\partial}{\partial x_i}\left[\rho u_i f_v - \rho D_{eff}\frac{\partial f_v}{\partial x_i}\right] = m_v''' \qquad (18)$$

Closure term for the species transport equation

$$D_{eff} = D_{mol} + \frac{\nu_t}{Sc_t} \qquad (19)$$

$$Sc_t = 0.7 \qquad (20)$$

**Governing Equations for the Lagrangian phase (droplet properties and parameters subscripted by d)**

Equation of motion of droplet

$$m_d \frac{d\vec{u_d}}{dt} = \frac{\pi d_d^3}{6}(\rho_d - \rho)\vec{g} + \frac{C_d \rho \pi d_d^2}{8}|\vec{u_d} - \vec{u}|(\vec{u_d} - \vec{u}) + \vec{F_{lift}} \qquad (21)$$

Energy-conservation equation, droplet phase

$$m_d C_{p,d}\frac{dT_d}{dt} = h\pi d_d^2(T - T_d) + \frac{dm_d}{dt}h_{fg} \qquad (22)$$

Energy transport from Lagrangian to Eulerian phase

$$Q_d = \frac{\sum_{i=1}^{i=N}\left(\pi d_{d,i}^2 h(T - T_{d,i})\right)}{V_{cell}} \qquad (23)$$

Ranz-Marshall correlation for Nusselt number

$$Nu = \frac{h d_d}{k_t} = 2.0 + 0.6 Re_d^{0.5} Pr^{0.33}, \qquad Pr = \frac{\mu}{\rho \alpha} \qquad (24)$$

Drag coefficient

$$C_d = \max\left\{\frac{24}{Re_d}(1 + 0.15 Re_d^{0.687});\ 0.44\right\} \qquad (25)$$

Lift force

$$F_{lift} = \frac{2K\nu^{0.5}\rho d_{ij}}{\rho_d d_d (d_{ik} d_{kl})^{0.25}} (\vec{u} - \vec{u_d}) \qquad (26)$$

Droplet evaporation term

$$\frac{dm_d}{dt} = \pi d_d^2 m_{wl} k_{mt}\left(\frac{p_{sat}}{RT_d} - X\frac{p}{RT}\right) \qquad (27)$$

Ranz-Marshall correlation for Sherwood number

$$Sh = \frac{k_{mt} d_d}{D} = 2.0 + 0.6 Re_d^{0.5} Sc^{0.33} \qquad (28)$$

Droplet Reynold number

$$Re_d = \frac{d_d \rho |\vec{u_d} - \vec{u}|}{\mu} \qquad (29)$$

Schmidt number

$$Sc = \frac{\mu}{\rho D} \qquad (30)$$

Mass of nonvolatile component

$$m_d^s = Y_0^s m_d^0 \qquad (31)$$

Mass fraction of nonvolatile component

$$m_d^s = Y_0^s m_d^0 \qquad (31)$$

Mass fraction of volatile component

$$Y_d^l = 1 - Y_d^s \qquad (33)$$

Droplet density

$$\rho_d = \frac{1}{\left(\frac{Y_d^s}{\rho^s}\right) + \left(\frac{Y_d^l}{\rho^l}\right)} \qquad (34)$$

Droplet heat capacity

$$c_{p,d} = Y_d^s c_{p,s} + Y_d^l c_{p,l} \qquad (35)$$

Droplet diameter

$$d_d = \left(\frac{6 m_d}{\pi \rho_d}\right)^{1/3} \qquad (36)$$

**C. Initial and Boundary Conditions**

The top circular mounting is assigned different conditions corresponding to three different scenarios i.e. quiescent, axial exhaust and exhaust fan . Table 1 provided below elucidates this further. The two outlets provided at the lower portion of the side walls are defined as pressure outlets with atmospheric pressure. All the walls and the boundary of the passenger are treated as walls with no slip velocity boundary condition. The door remains closed in all the scenarios and hence in all these scenarios, the door is modelled as a wall. By conducting numerical simulations, it has been inferred that the initiation time of the cough and air flow development in the domain has significant impact on the droplet kinematics. To eliminate this above stated bias and to make the situation more representative and generic, the cough is ejected in stages. The complete coughing phenomenon occurs in four stages. As depicted in Fig. 2 , the coughing phenomenon commences at 1s and terminates at 4.12s. Each single cough occurs for 0.12s, injecting 1008 droplets of mass 7.7 mg with a velocity of 8.5 m/s normal to the mouth surface[29]. In each stream, the initial size distribution of the cough droplets follows the well-known Rosin-Rammler distribution or the Weibull distribution with a scale factor of 80 µm and shape factor of 8 [29].The passenger, assumed to be a symptomatic COVID infected patient, has a comparatively higher body temperature of 38.4°C. The continuous periodic inhalation and exhalation of the passenger has also been taken into account. The ambient pressure inside the elevator is one atmospheric pressure, 101325 Pa. The elevator is subjected to two different ambiences, 30°C, 50% R.H. ( Hot dry) and 10°C, 90% R.H. (Cold humid). The temperature of the air ejected out of the mouth during exhalation is assumed to be at the body temperature and its relative humidity is assumed to be 100%[29]. The injected cough droplets are also at the body temperature. The cough droplets are considered as a mixture of NaCl (salt) and water with initial mass fractions of 0.01 NaCl (solid) and 0.99 water (liquid)[45].

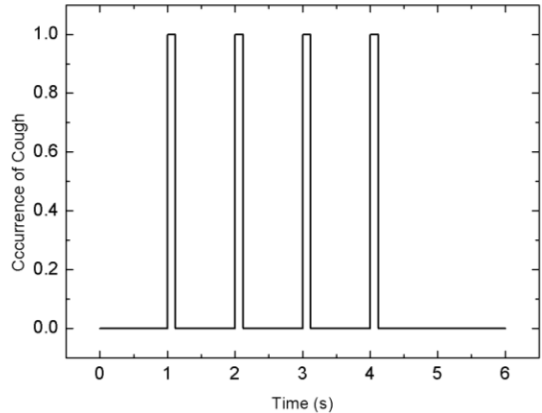

**Fig. 2 : The entire coughing phenomenon**

**D. Numerical Method**

The OpenFOAM solver "reactingParcelFoam", with necessary modifications to successfully implement the salt model as discussed earlier was employed to solve all the required partial differential equations. It is very important to state that all the thermophysical properties of the Eulerian and Lagrangian phases are functions of temperature. The Eulerian phase has been modelled as an ideal gas for its equation of state, and its transport is modelled using Sutherland's law[46] for its viscosity based on the kinetic theory of gases, which is suitable for non-reacting gases. Finite volume methods have been employed to discretize the Eulerian phase. Second-order schemes have been employed for both space and time operators. Semi-implicit numerical schemes of second order have been employed for Lagrangian phase discretization. The grid independence study has been performed which to improve the readability of the paper has been provided in Appendix section.

**III. Validation**

**I. Validation of Droplet Evaporation Model**

Many previous studies have been conducted but the authors did not take into consideration the effect of soluble components present in the cough droplets. In reality, cough droplets are not pure water and contain dissolved salts (like NaCl) in certain proportions. This presence of salts in cough droplets affects the droplet characteristics in several interconnected ways, as discussed in the Introduction section. In our study, an attempt is being made to make the simulations more realistic by considering the effect of salt solution in cough droplets i.e. by including the salt model of droplets. Our model is tested against the reported experimental result of change of diameter of an acoustically levitated 1 wt% salt (NaCl) laden droplet with time of Basu et al.[45] For this validation study, only 1 droplet (NaCl 1% by wt., $H_2O$ 99% by wt.) is injected with initial size of 600 µm at the center of a domain. The quiescent condition is modelled appropriately by taking the domain size much larger than the droplet diameter and by assigning the internal field as well as the boundary fields of the entire computational domain a zero velocity. A temperature of 30℃ or 303K and a relative humidity of 50% is used as reported by Basu et al.[45] In order to model the levitating droplet, the droplet is injected with zero initial velocity and no force (gravitational, buoyancy or sphere drag) is applied on the droplet, thus keeping it suspended in the domain indefinitely. The droplet diameter reduces continuously owing to its evaporation and the change in diameter (D) with time is noted. Figure 3(a) compares experimental data (Basu et al.[45]) of the temporal history of the instantaneous normalized droplet diameter ($D/D_0$; $D_o$ initial diameter) with our numerically predicted results. The numerically predicted results are in a reasonably good agreement with the experimental observations of Basu et al.[45], as can be seen from the two graphs in Fig. 3(a) . Hence our newly developed droplet salt model is validated.

**II. CFD Model Validation**

Before proceeding with the numerical model for our actual study, a quantitative validation of the droplet size distribution is done against the DNS data of size distribution reported by Ng et al.[13]. The validation data used are for temperature 30℃ and R.H. 90% , of Ng et al.[13]. For this validation study, the geometry and all the initial and boundary conditions are that of Ng et al.[13]. The droplet size distribution as predicted by our numerical (CFD) model is compared with the size distribution results of Ng et al[13]. Figure 3(b) compares the droplet size distribution at t = 0.6s. The match between the droplet size distributions is reasonably good. Thus, our numerical (CFD) model is validated and hence, our numerical (CFD) model along with our newly developed salt model can be used subsequently in our actual study.

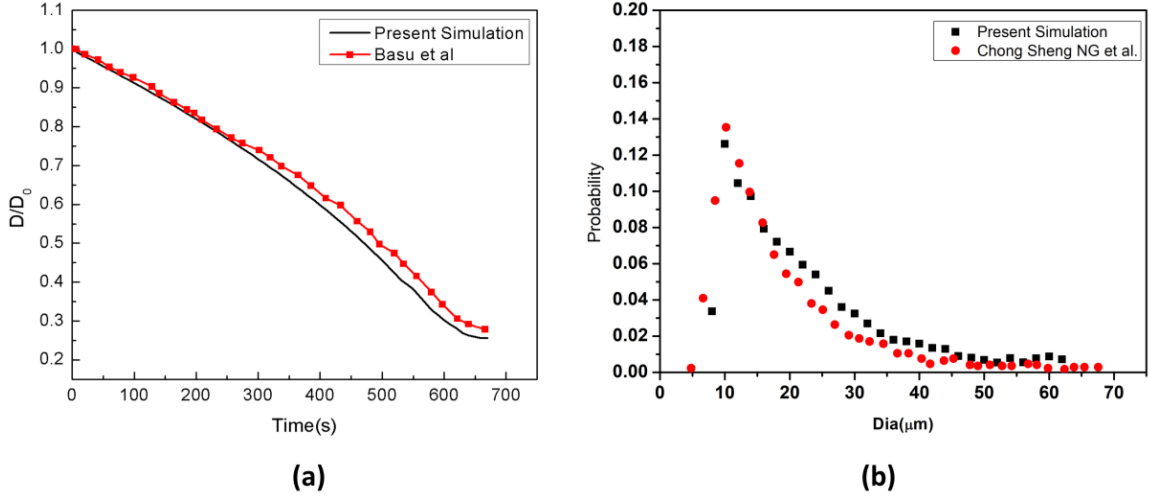

**Fig. 3 : (a) Validation of droplet evaporation model including crystallisation with literature data
(b) Validation of Droplet size distribution with literature data at t = 0.6s**

| Ambiences | Table 1: Table of Scenarios, *ĵ indicates positive y direction. | | | |
|---|---|---|---|---|
| | Scenario | Objective | Axial Velocity of top circular mounting (m/s) | Angular Velocity of top circular mounting (rpm) |
| Hot dry (30°C,50% R.H.) | 1 | Effect of quiescent environment | 0 | 0 |
| | 2 | Effect of Axial Exhaust | 0.56ĵ | 0 |
| | 3 | Effect of Exhaust Fan | 0.56ĵ | 2000ĵ |
| Cold humid (10°C,90% R.H.) | 1 | Effect of quiescent environment | 0 | 0 |
| | 2 | Effect of Axial Exhaust | 0.56ĵ | 0 |
| | 3 | Effect of Exhaust Fan | 0.56ĵ | 2000ĵ |

**IV. RESULTS AND DISCUSSIONS**

In this study, two different ambient conditions and three different types of ventilation scenarios in each of the ambient conditions, (as summarized in Table 1), have been studied.

In the first ambient condition, a hot dry ambience having temperature of 30°C and relative humidity of 50% has been investigated that closely matches the climatic condition of places like Mumbai, Delhi, New York, Melbourne, etc. in summer. Besides this, in the second ambient condition, an ambience of cold humid having a temperature of 10°C and relative humidity of 90% found sometimes in places like London, Amsterdam, Berlin, etc. has also been investigated to study the effect of change in environment on droplet transmission. In both the ambient conditions, three different scenarios are studied.

In scenario 1, the effect of a quiescent environment (i.e. no airflow or ventilation condition) is studied. In scenario 2, the effect of an outward (exhaust) axial jet ventilation condition is studied. In scenario 3, the effect of an exhaust fan ventilation condition is studied. These scenarios have been studied and quantified to progressively move towards a ventilation system that minimizes the chances of getting infected for a passenger/passengers if they had been travelling in the elevator (or is waiting outside to enter the elevator) with the above mentioned infected passenger considered in the study.

In Scenario 1, the effect of a quiescent environment on droplet dynamics and heat transfer characteristics have been studied whereas scenarios 2-3 investigate the effect of forced circulation on the same.

The scenarios 1-3 have been investigated for 10 seconds, the average time taken by an elevator to traverse 10 floors, considering average height of a floor as 3 meters (complying with National Building Code of India, 2005)[47] and mean elevator speed as 3 m/s.

### A. Transport and Evaporation of Droplets

#### 1. Scenario I

In this scenario, the top mounting is treated as a wall, with no airflow interaction with the domain, thus making a quiescent environment prevail inside the elevator. For the hot dry case, the droplets do not reach the floor within the stipulated elevator travel time of 10 seconds, as established by Fig. 4(a)(Multimedia View) . It is important to note that droplets do not directly head towards the floor rather they get entrapped in the turbulence induced both by the cough and the continuous inhalation and exhalation of the passenger, and they get spread across the elevator. The absence of any continuous draft of air in the domain slows down the process of sticking or escaping of the droplets, hence majority of them remain suspended in the air. The suspended droplets evaporate continuously to decrease in size as visible from the continuous shrinking size range, depicted by the diameter distribution of the suspended droplets at various time instances in Fig. 5 (a-b) . A probability distribution plot based on the initially injected droplet count has been used to represent the size distribution in Fig. 5 . Figure 5 also depicts the probability of droplet nuclei being formed. Droplet nuclei are the droplets from which the volatile liquid component has completely evaporated and is remaining only with the nonvolatile salt component. Due to the initial droplet size distribution following a Rosin-Rammler distribution, we have droplet nuclei of different sizes (Fig. 5(a-b) ) and also it is possible to have a droplet and a droplet nuclei of same size (Fig. 5(a) ). In this context, it is notable that out of a droplet and a droplet nuclei of same size, the droplet nuclei is more harmful as it has surely been inherited from an initially larger size droplet, thus exhibiting a very high viral load[11]. Also, the droplet nuclei basically consist of only the nonvolatile component, which in turn contains the pathogen. A significant number of suspended droplets evaporate to form droplet nuclei having size mostly in the range of 10-20 µm after 10s as shown in Fig. 5(a) . Droplet nuclei are only found in the hot dry ambience and not in the cold humid ambience as in the later there is negligible evaporation thus preventing the formation of nuclei.

Contrasting results are obtained for the cold humid ambience as shown in Fig. 4(b) as well as from the size distribution at various instants of elevator travel time as can been seen from Fig. 5(b) . Due to their relatively larger mass as compared to the hot dry condition owing to negligible evaporation, the gravity force dominates over the injected droplets which ultimately cause the droplets to descend and reach the elevator within the stipulated time of 10s as can be seen in Fig. 4(b) . Moreover, the relatively larger diameter due to negligible evaporation, as compared to the hot dry condition produces a larger drag force on the droplets which prevents the spread of droplet in the elevator and produces an orderly downward motion. Although a few droplets are initially trapped in the turbulent puff of coughing and spread wayward, but owing to the gravity effect and larger drag force, they quickly settle down.

#### 2. Scenario II

In this scenario the effect of forced circulation in the domain (for both the environmental condition), on droplet dispersion is investigated. Air is drawn out of the domain and the top mounting is modelled as an axial exhaust jet. As can be seen from the droplet distribution (Fig. 4 (c, d))(Multimedia View) , for the both the environmental conditions the elevator space (mostly the upper portion) is filled with many suspended particles. This is due to the fact that the first two streams of cough particles move down instead of moving up, while the third and fourth ones move up. This difference occurs because the flow has not developed during the time of injection of first two streams as depicted by the velocity contours at plane AB (plane AB shown in Fig. 6 ) of different time instances in Fig. 8 , where the development of velocity in the flow field along with the droplets' positions have been depicted for the hot dry ambience. Similar flow pattern is observed for the cold humid ambience. As the flow takes sufficient time to develop as reconfirmed by the temperature contoured velocity vector plots of Fig. 7 , for hot dry ambience at plane AB (Fig. 6 ), the initially injected particles do not receive sufficient drag force to move upwards hence descends (due to gravity), but as the flow becomes fully developed, the downward motion of the particles get arrested and these particles remain suspended in the domain for an extended period of time. As air is drawn out in this scenario, a significant percentage of droplets escape out of the domain. For the hot dry condition, the suspended droplets evaporate continuously to decrease in size as indicated by the continuous decreasing diameter range of the diameter distributions of the suspended droplets, depicted at various time steps by Fig. 5(c) while majority of the suspended droplets turn into droplet nuclei (having size in the range of 10-20 µm) after 10s. Whereas for the cold humid condition it is found just like the previous scenario, there is negligible evaporation of droplets preventing the generation of droplet nuclei, Fig. 5(d) .

**3. Scenario III**

In this scenario the top mounting is modelled as an exhaust fan. Here the fan rotates with a r.p.m of 2000 to suck air out of the domain. Unlike scenario 2, where the top mounting was modelled as an axial exhaust jet the rotational effect of fan has also been considered. In contrast with the previous scenario, the flow develops rather quickly in the domain (especially near the man's mouth) to develop a sufficient strong drag force, for both the ambient conditions as depicted by the temperature contoured velocity vector plots in Fig. 9 . This is further elucidated by the velocity contour plots of Fig. 10 . Hence, the particles move upwards immediately upon injection due to the more enhanced drag force exerted by air on the particles, as can be seen from the droplet dispersion transience depicted in Fig. 4 (e-f)(Multimedia View) . The circulation brought about by the rotational effect, increases the dispersion in droplet kinematics due to the additional turbulence (created by the rotating component of the fan), due to which the droplets rise up and a significant amount of them gets deposited at the roof (top wall) of the elevator, quite contrary to the previous scenario of axial exhaust jet. As can be understood from the droplet distribution in Fig. 4 (e-f) (Multimedia View), for the hot dry ambience, after 5.5 s, none of the droplets remain below the height of the passenger whereas for the cold humid ambience, the elevator becomes safe but it takes 7.48 s. Hence the domain becomes completely safe from these time instants for both the ambient conditions. Since a negligible percentage of injected droplets remain suspended in the domain at all time instants, the size distribution has not been investigated.

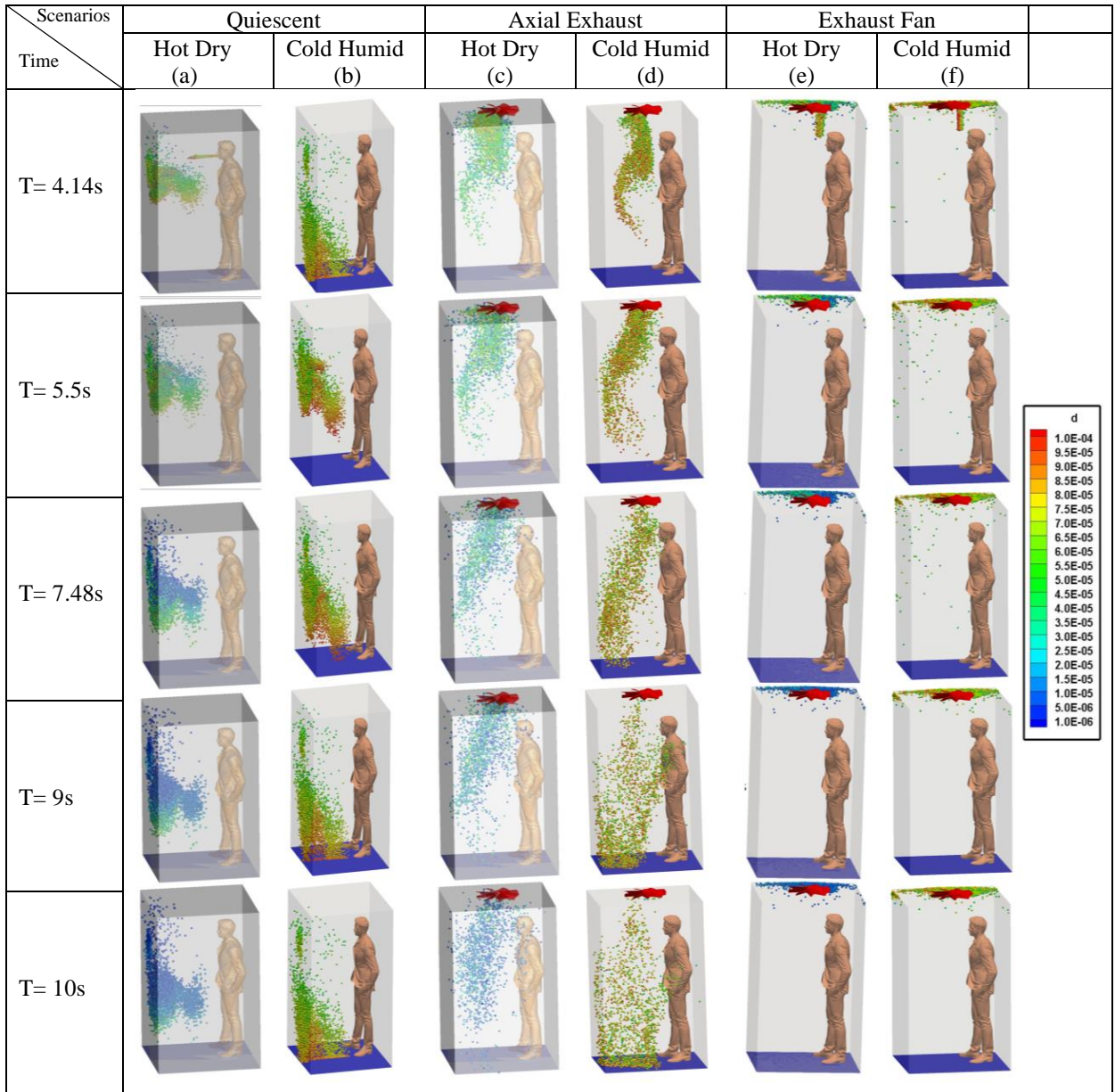

**Fig. 4 :(a)-(f) Droplet dispersion in the domain at various time instants for all the scenarios of both the ambiences(Multimedia View).**

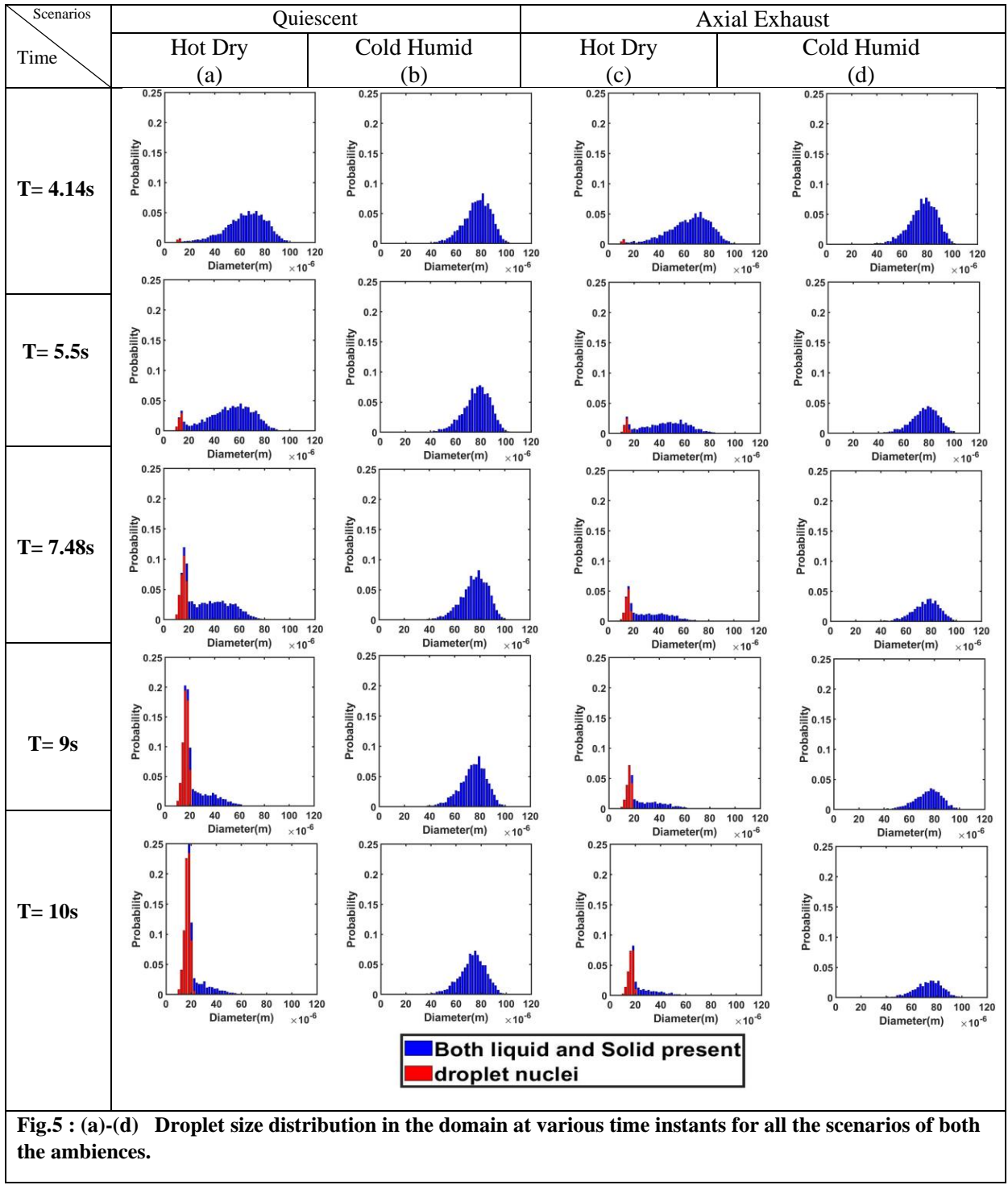

Fig.5 : (a)-(d)   Droplet size distribution in the domain at various time instants for all the scenarios of both the ambiences.

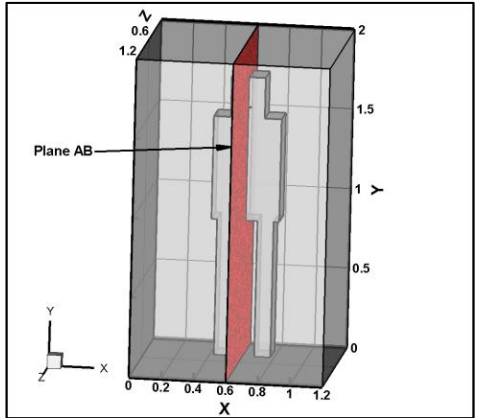

**Fig. 6 : Showing cross-sectional plane AB .**

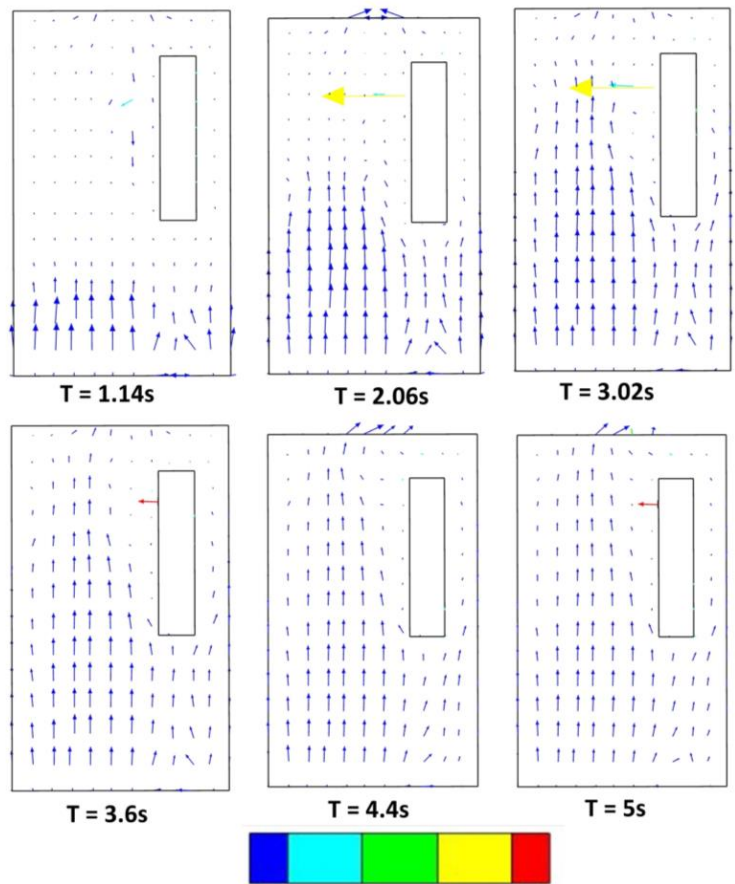

**Fig. 7 : Temperature contoured velocity vector plots at plane AB (of Fig. 6 ) at different time instances, of scenario 2 for the hot dry ambience.**

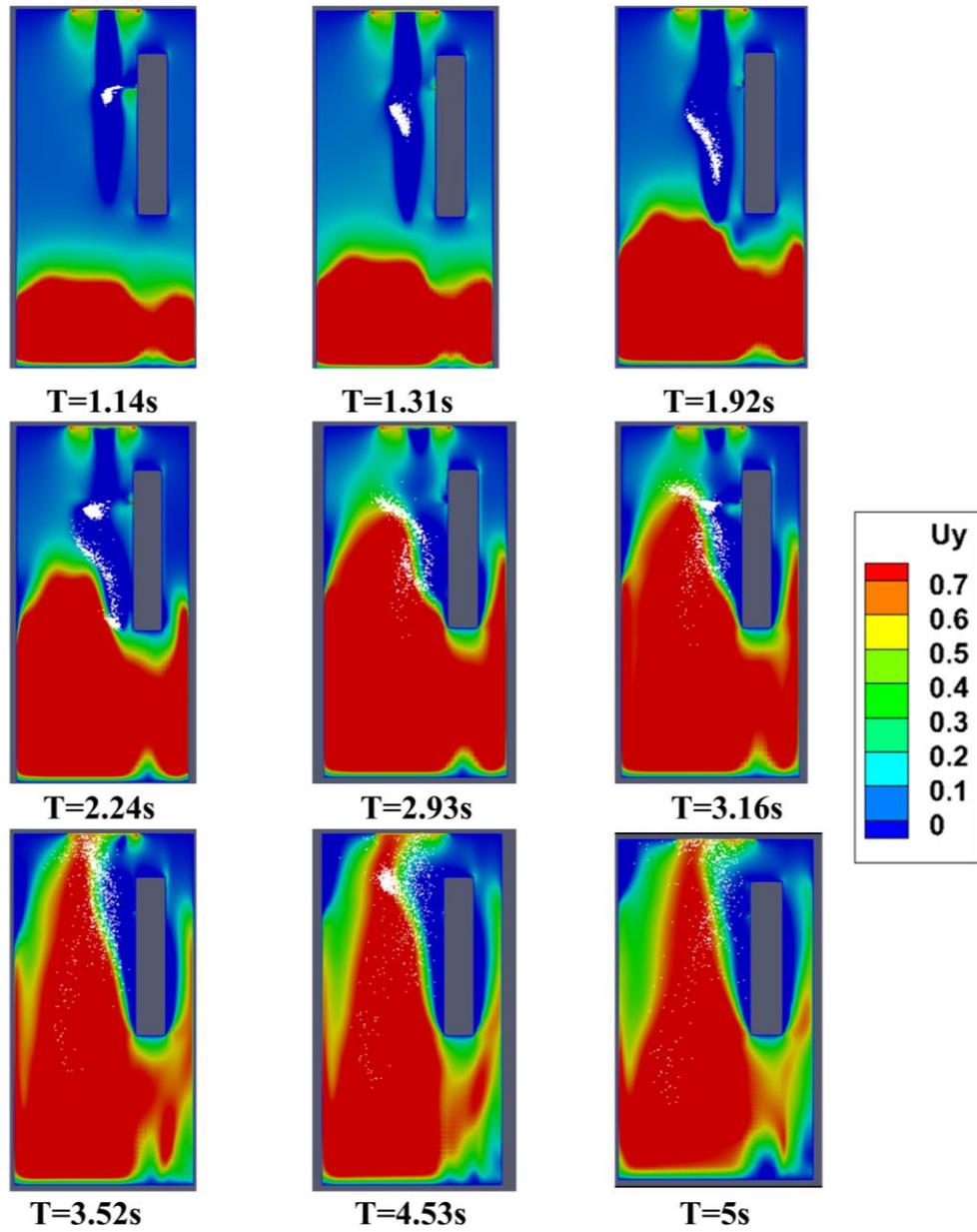

**Fig. 8 :** Velocity contours at plane AB (of Fig. 6 ) along with droplets, at various instances of Scenario 2 for the hot dry ambience.

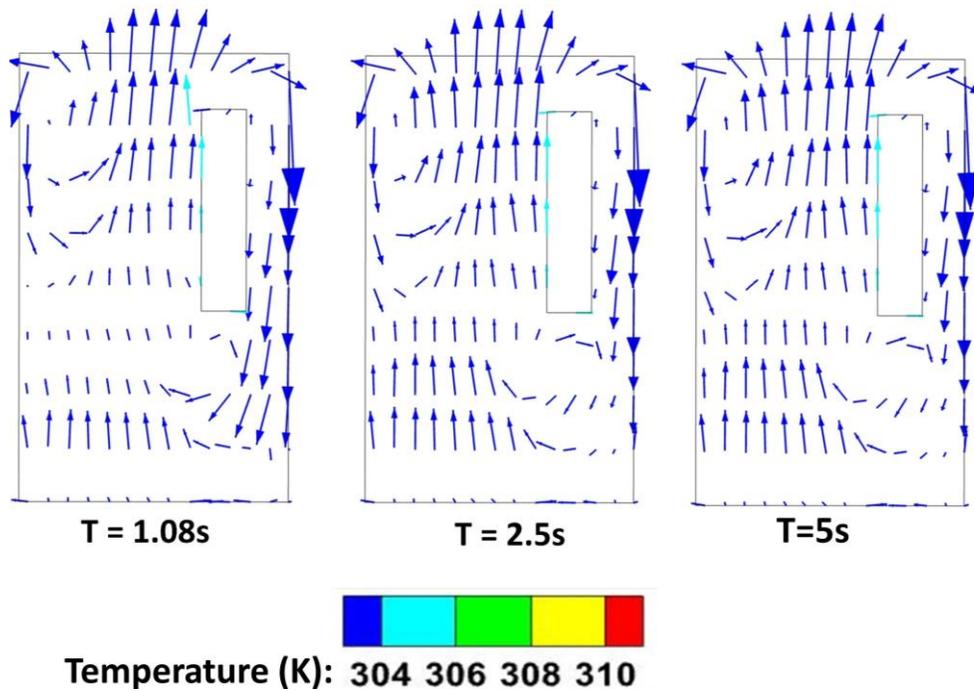

**Fig. 9 :** Temperature contoured velocity vector plots at plane AB (of Fig. 6 ) at different time instances, of scenario 3 for the hot dry ambience.

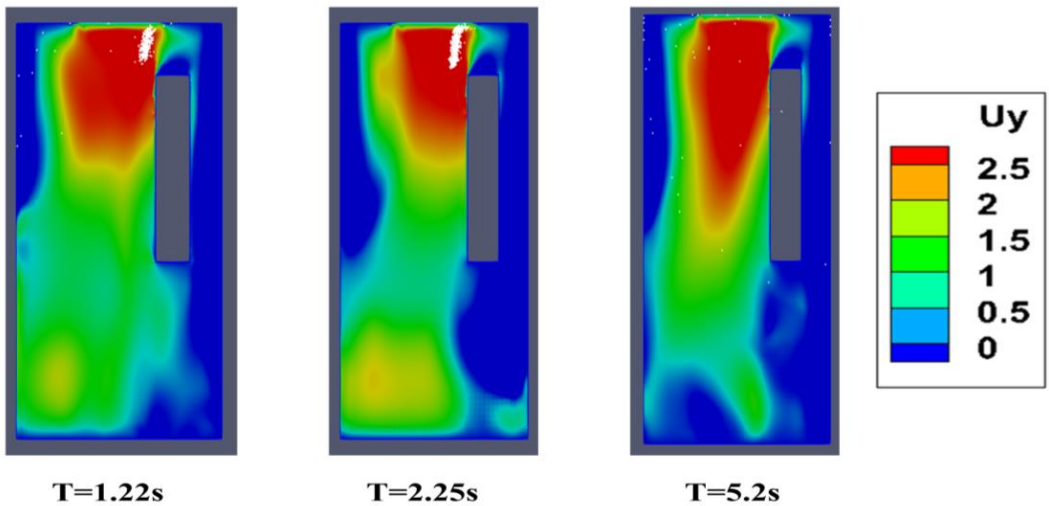

**Fig. 10 :** Velocity contours at plane AB (of Fig. 6 ) along with droplets, at various instances of Scenario 3 for the hot dry ambience.

## B. Implications for Spread of Disease

It is a critically acclaimed fact that the two main modes of spread of coronavirus are by touch and inhalation of particles having significant viral load. So, it is critical that we investigate how the dispersed droplets in the domain might come in contact with a person and obtain a better understanding on this aspect. At first it is important to quantify how the droplets disperse in the domain by enumerating the percentage of injected droplets that remains suspended in the domain, that escapes from the domain and the percentage that gets deposited on the elevator surfaces. The height range 0.8m to 1.8m, (average height from a person's waist to head) is identified as the risky height zone as droplets (either suspended or deposited on surfaces) in this zone will be most perilous to any other person travelling in the elevator as these droplets may be directly inhaled by the other passengers. Figure 11 shows the percentage of injected droplets that remains suspended in the domain (within and outside the risky height zone), the percentage that escapes and the percentage of droplets that gets deposited within and outside the risky height zone on the elevator surfaces, for the various scenarios. A parameter called Risk Factor is defined as the time averaged percentage of injected particles that remains suspended in the domain within the risky height zone over an inhalation time period and this parameter is an indicator of the probability of an exposed passenger to get infected. The period of averaging being a typical inhalation cycle of 1s and within this time span of 1s, the values of percentage of injected droplets in the risky height zone at 5 different instants has been computed. Our defined Risk Factor is the average of these values. In the bottom-right corner of Fig. 12 , 5 instantaneous values of percentage of droplets suspended in the risky height zone (in black) and the average of these instantaneous values representing the defined Risk Factor for the period of 6-7s (in red) has been depicted (for one particular representative of cold humid quiescent scenario) in order to elucidate the above mentioned averaging process. This risk factors has been depicted at different time instants for the different scenarios in Fig. 12 . It is understood from Fig. 11 and Fig. 12 , that in a quiescent domain with hot dry condition, where there is no continuous draft of air, almost all the droplets remain suspended in the domain and only a few percentages get deposited on surfaces. Also, a fair percentage of suspended droplets remain in the risky height zone at all times (Fig. 12 ), thus producing the highest risk factor of all the scenarios, rendering a bleak future to the elevator and its other passengers. Whereas for the cold humid ambience the droplets settle below the risky height zone owing to their larger masses at latter stages of stipulated time bringing down the risk factor significantly below that of hot dry ambience as shown in Fig. 12 .

The introduction of a forced circulation of air in the form of axial exhaust jet alleviates the problem for hot dry ambience and produces a significant change in droplet dynamics, as can be seen from Fig. 11 , by significantly reducing the percentage of droplets that remains suspended in the domain and increasing the percentage of droplets that escapes out of the domain. The first and second streams of droplets move downwards initially due to gravity force ( which is not balanced by the weak drag force) thus ultimately contributing to producing a high-risk situation. As the flow gets developed in the domain the third and fourth stream of droplets go up whereas the droplets from the first two streams remain suspended in the domain due to the stronger drag force now available due to complete development of the flow.

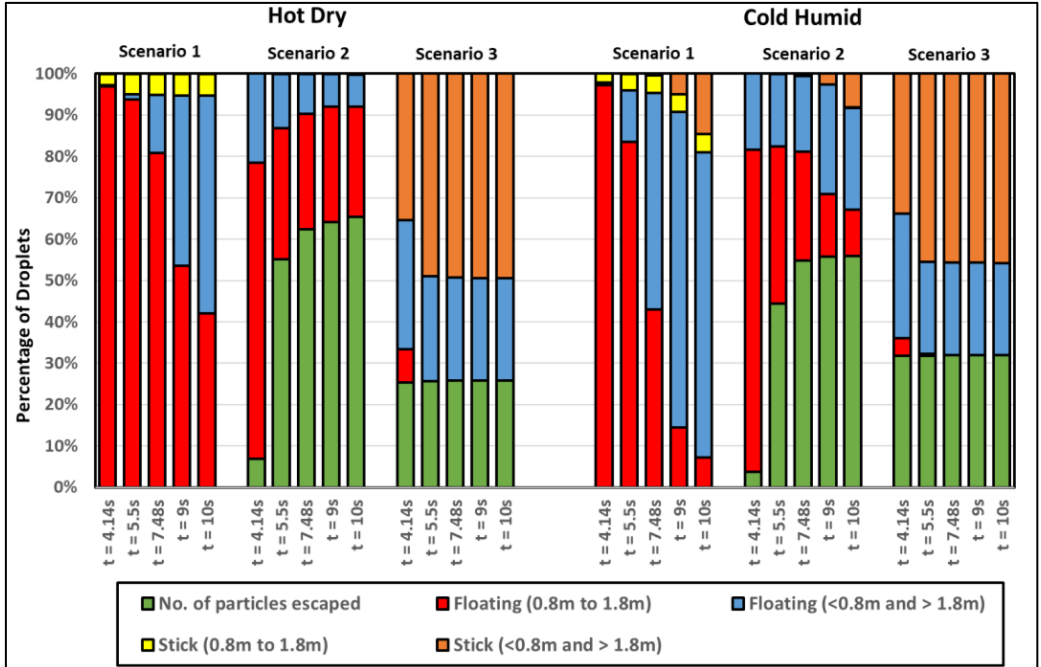

**Fig. 11 : Showing droplet fate at various time instants for all the scenarios of both the ambiences.**

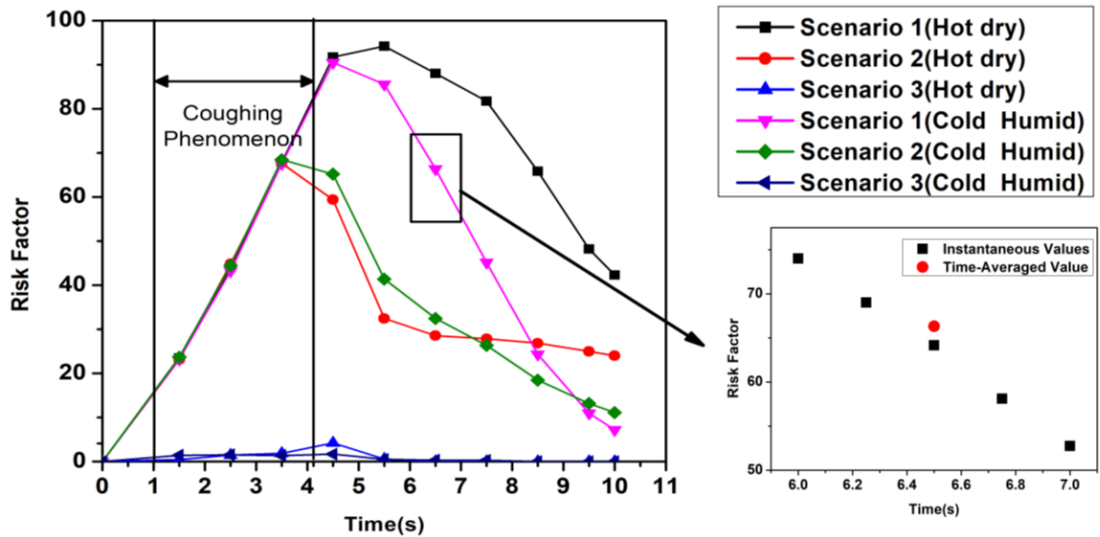

**Fig. 12 : Showing risk factor for various scenarios along with the elucidation of the time averaging process for the computation of risk factor in the bottom right corner.**

Although this situation produces a significant amount of risk in the domain, but as a significant percentage of the third and fourth stream droplets escapes out of the domain, the situation improves as compared to the quiescent scenario producing a lower risk as compared to the quiescent scenario. On the contrary, for the cold humid ambience, although the axial exhaust scenario initially ameliorates the situation as compared to the quiescent scenario, at later stages the risk associated with the quiescent scenario falls below that of the axial exhaust as depicted in Fig. 12 . In this case of axial exhaust jet as discussed, the flow takes some time to develop and the droplets remain suspended in the domain thus producing a higher risk at later stages as compared to the quiescent scenario where the droplets ultimately settle down below the risky height zone owing to their larger diameter due to reasons discussed earlier. Additionally, it is found from Fig. 4(d) that in the cold humid ambience a significant percentage of the initially injected large droplets quickly fall below the risky height zone owing to larger masses and negligible evaporation. This causes the risk factor for this scenario in cold humid ambience to be below the hot dry ambience as can be understood from Fig. 12 .

In the third scenario where the top mounting has been modelled as the exhaust fan taking into account its rotation. This ventilation condition is the best solution for ensuring minimum risk as compared to the others. The circulation brought about by the rotational effect, increases the dispersion in droplet kinematics due to the additional turbulence (created by the rotating component of the fan) as can be understood from the velocity contour plots of Fig. 10 , due to which the droplets rise up and a significant amount of them gets deposited at the roof (top wall) of the elevator, quite contrary to the previous scenario of axial exhaust jet. The risk for this scenario remains approximately equal and significantly low for both the ambiences at all instants. Furthermore, it is found that after 5.5 seconds for hot dry scenario and 7.48 seconds for the cold humid ambience there are no droplets in the risky zone hence no there is no risk of being infected. More time required for the cold humid ambience is understandable due to larger masses of droplets owing to negligible evaporation. Thus, it can be concluded that the exhaust fan ventilation condition is the best solution for minimizing risk of infection in the elevator. As well as it is understood for all other ventilation scenarios the hot dry ambience poses a higher risk as compared to the cold humid ambience.

Till now we have emphatically established the fact that out of all the ventilation scenarios, scenario 3 is the best-case scenario as after some time there remains no droplet in the risky height zone making the domain completely safe and the risk falls down to zero. But it is also important to note that if any of the other scenarios are being used as ventilation conditions in an elevator, the safe radial distances to be maintained from the mouth of the infected passenger to ensure maximum safety in the elevator must also be investigated. Hence, firstly height of each suspended droplet in the domain has been tracked at the end of 10s as shown in Fig. 13(a-f) , and secondly, among these suspended droplets, the radial distance of those droplets suspended in the risky height zone has been tracked, as depicted in Fig. 14(a-d) to ascertain the minimum radial distance to be maintained in all the cases to avoid as much as possible, coming in contact with any droplet in the risky height zone. The safe distances for all scenarios for both ambiances are enumerated in table 2. For both, the ambience quiescent scenario requires maintaining the maximum radial distance from the passenger's mouth to get rid of the palpable danger pervading the elevator domain. In fact, for the cold humid ambiance, droplets are present at the farthest possible distance from the passenger's mouth. Except scenario 3, where the domain gets completely safe quickly, the cold humid ambience requires maintain greater radial distance for complete safety as compared to hot dry ambience.

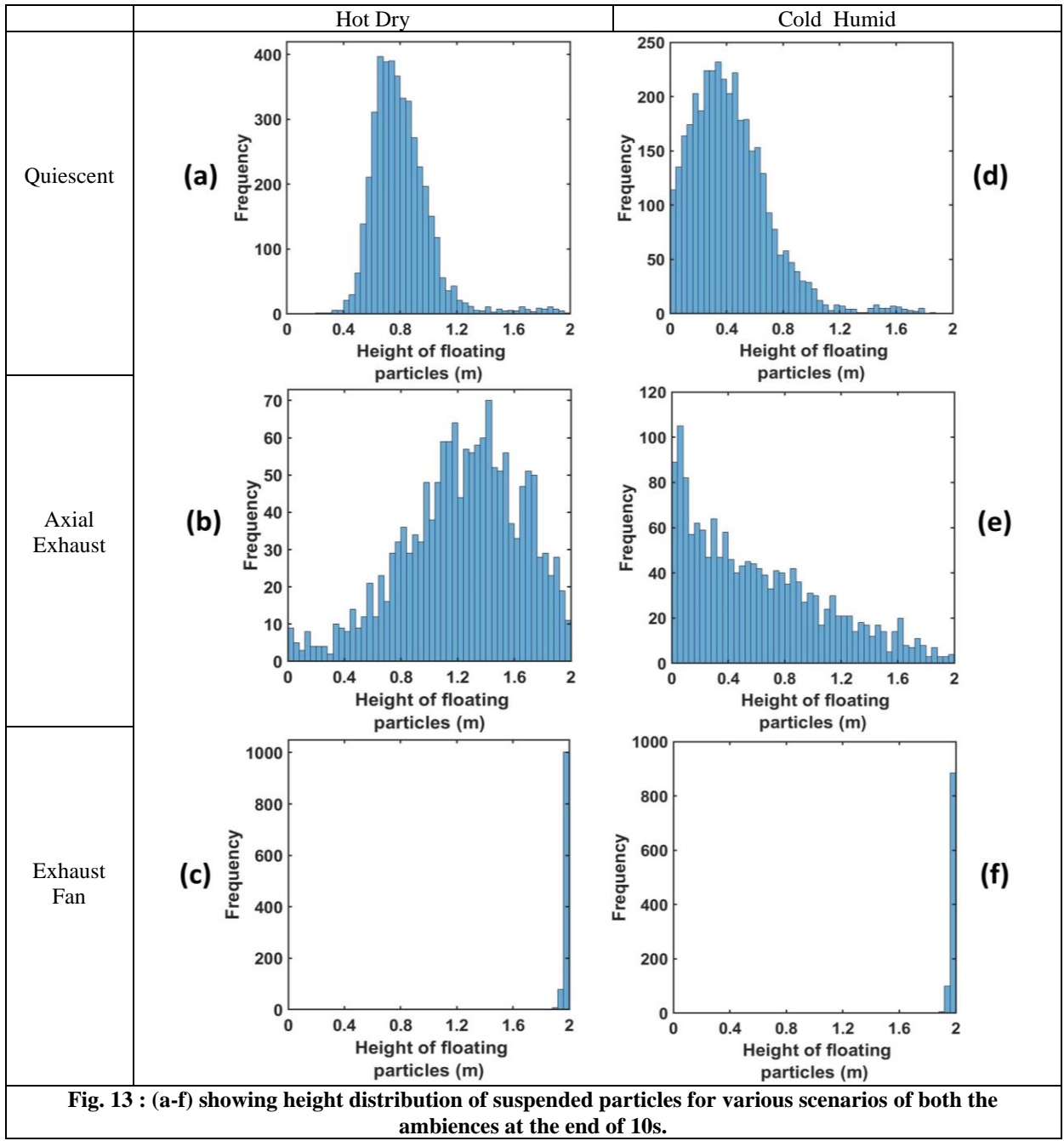

**Fig. 13 : (a-f) showing height distribution of suspended particles for various scenarios of both the ambiences at the end of 10s.**

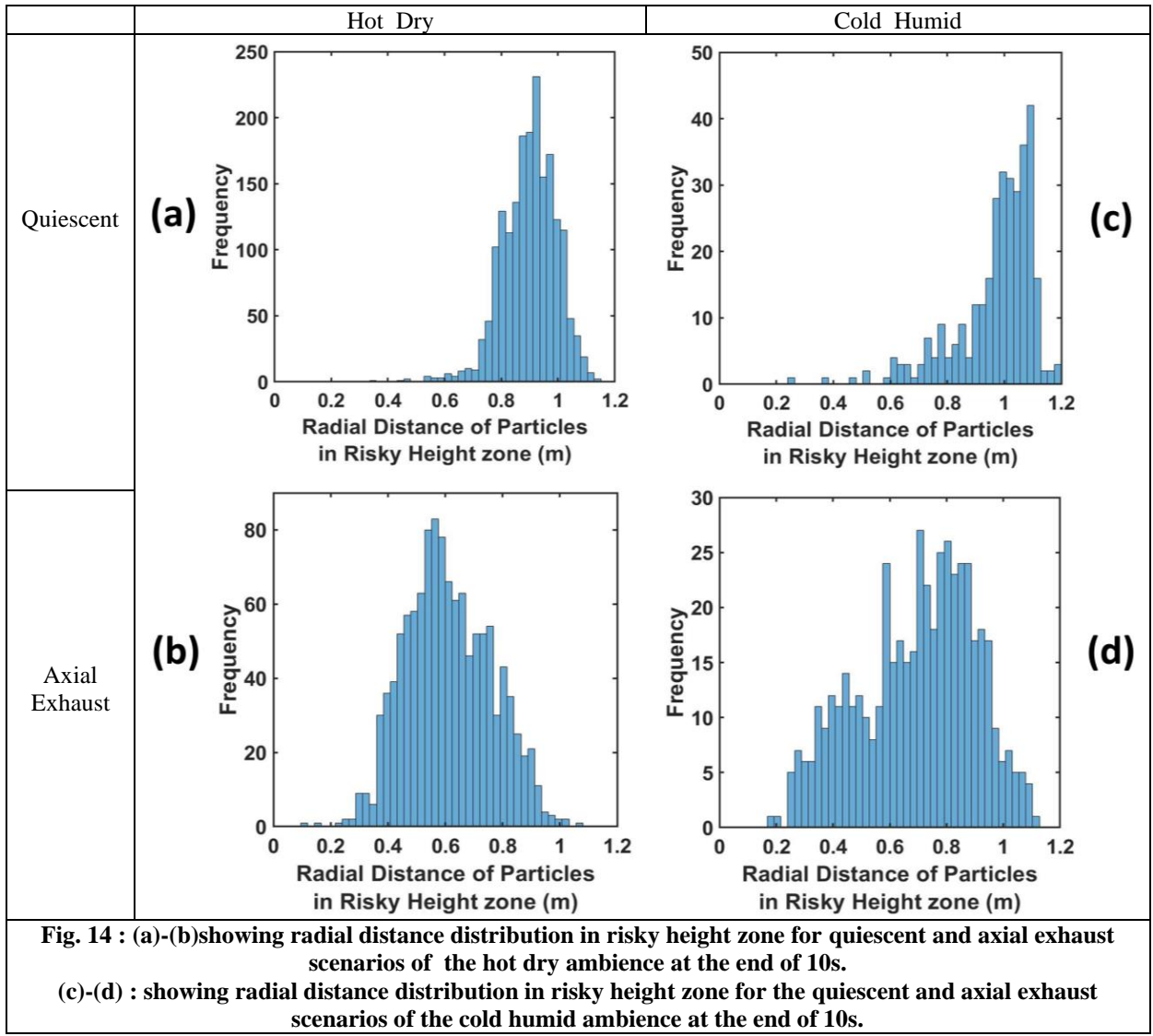

**Fig. 14 : (a)-(b) showing radial distance distribution in risky height zone for quiescent and axial exhaust scenarios of the hot dry ambience at the end of 10s.**
**(c)-(d) : showing radial distance distribution in risky height zone for the quiescent and axial exhaust scenarios of the cold humid ambience at the end of 10s.**

Another important fact that was investigated was the rate of evaporation of the suspended droplets in the domain – indicated by the rate of formation of droplet nuclei from the suspended droplets. This is only applicable for the hot dry ambience (as there is negligible evaporation in the cold humid scenario. It is important to understand that these droplets will continuously evaporate to form very small-sized particles that will remain suspended in air for a very long duration and will have a high chance of infecting any other person travelling in the elevator. Figure 15 shown below depicts the percentage of suspended droplets that have fully evaporated to form droplet nuclei at different time instances for various scenarios. It can be seen for the forced circulation scenario of exhaust fan, the evaporation rate is higher than that of quiescent scenario (scenario 1). This is because in the scenario of exhaust fan, the increased velocity of the droplets increases the evaporation rate owing to the increased Sherwood number for the droplets. Moreover, there is an increased dispersion of droplets in the elevator due to the additional turbulence brought about by the rotation of the fan. But in the case of axial exhaust jet (although it is a forced circulation scenario), the evaporation rate is slower as compared to the other scenarios because it takes significant amount of time for the flow to develop and droplet cloud from the first two streams travel together in a cluster remaining condensed (and suspended) rather than spreading across the domain hence residing in areas having locally higher concentration of water vapour, thereby slowing the evaporation rate. Still we can say that a significant fraction of suspended droplets gets fully evaporated to droplet nuclei at a fairly fast rate. This high evaporation rate is a matter of concern, as a

Submitted to arXiv ( Accepted in Physics of Fluids, Flow & Virus)

significant percentage of injected droplets quickly evaporates to form viruosols, i.e. particles having very high viral load and diameter less than 20µm which remain suspended in the domain for a long period of time[49]. These virusols are the most dangerous of all the droplets. Because of their size (dia<20µm), they have the highest penetration in human lungs[50]. A comparison of the percentage of injected droplets that have formed suspended virusols in the risky height zone at different time instants is shown in Fig. 16  below for the hot dry ambience. The virusol count is negligible (close to zero) for cold humid condition owing to negligible evaporation. Scenario 1 (hot dry ambience) has the highest concentration of virusols suspended in the risky height zone (Fig. 16), thus aggravating the threat of infection inside the elevator. Figure 17(a-d)  portray the size ranges of the droplets suspended in the risky height zone after 10s, thus giving an idea of the viral load residing in this regime and it also tracks the droplets' radial locations and their concentration in a single plot, for scenarios 1 and 2 (Since in Scenario 3 no droplets remain after 5.5s or 7.48s in the domain for both the ambiences). Figure 18(a-b))  extracts out the virusols (for col dry ambience) and shows separately the radial concentration of these virusols suspended in the risky height zone i.e. the number of virusols suspended at different radial locations from the mouth and thus give an idea of the most critical radial distance i.e. the radial distance having the highest concentration of malicious virusols. The most critical (or the most dangerous) radial distance for the different scenarios (of hot dry ambience) is enumerated in Table 3. This most critical radial distance must always be eluded by the other passengers to somewhat avert the chances of infection.

| Table 2: Table showing safe radial distances for various scenarios. | | |
|---|---|---|
| **Ambiences** | **Scenarios** | **Distance (m)** |
| Hot dry | Scenario 1 | 1.1312 |
|  | Scenario 2 | 1.0682 |
|  | Scenario 3 | Entire Domain is Safe |
| Cold humid | Scenario 1 | 1.2 |
|  | Scenario 2 | 1.13 |
|  | Scenario 3 | Entire Domain is Safe |

.

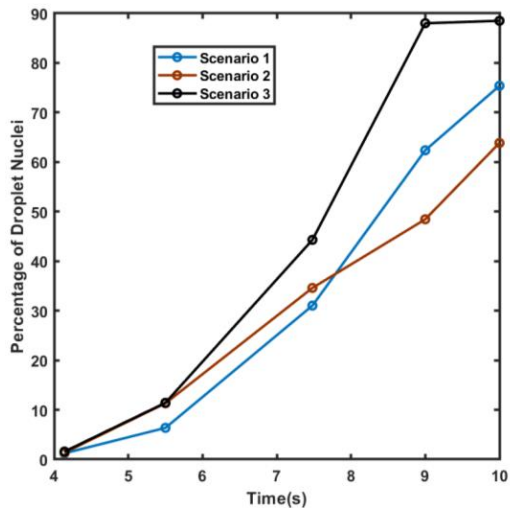 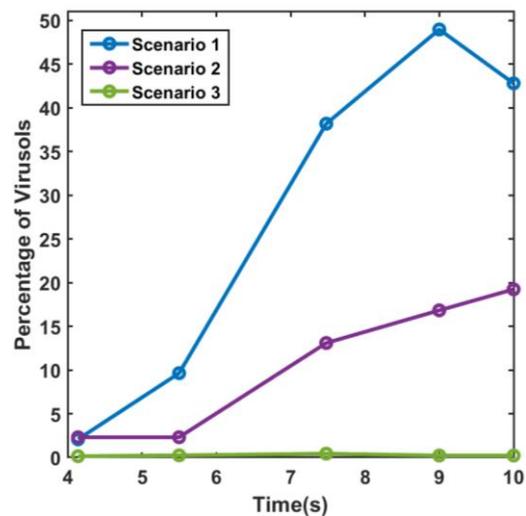

**Fig. 15 : Percentage of suspended droplets that have fully evaporated to droplet nuclei at different time instances for various scenarios for the hot dry ambience.**

**Fig. 16 : Percentage of suspended virusols in the risky height zone at different time instances for various scenarios for the hot dry ambience.**

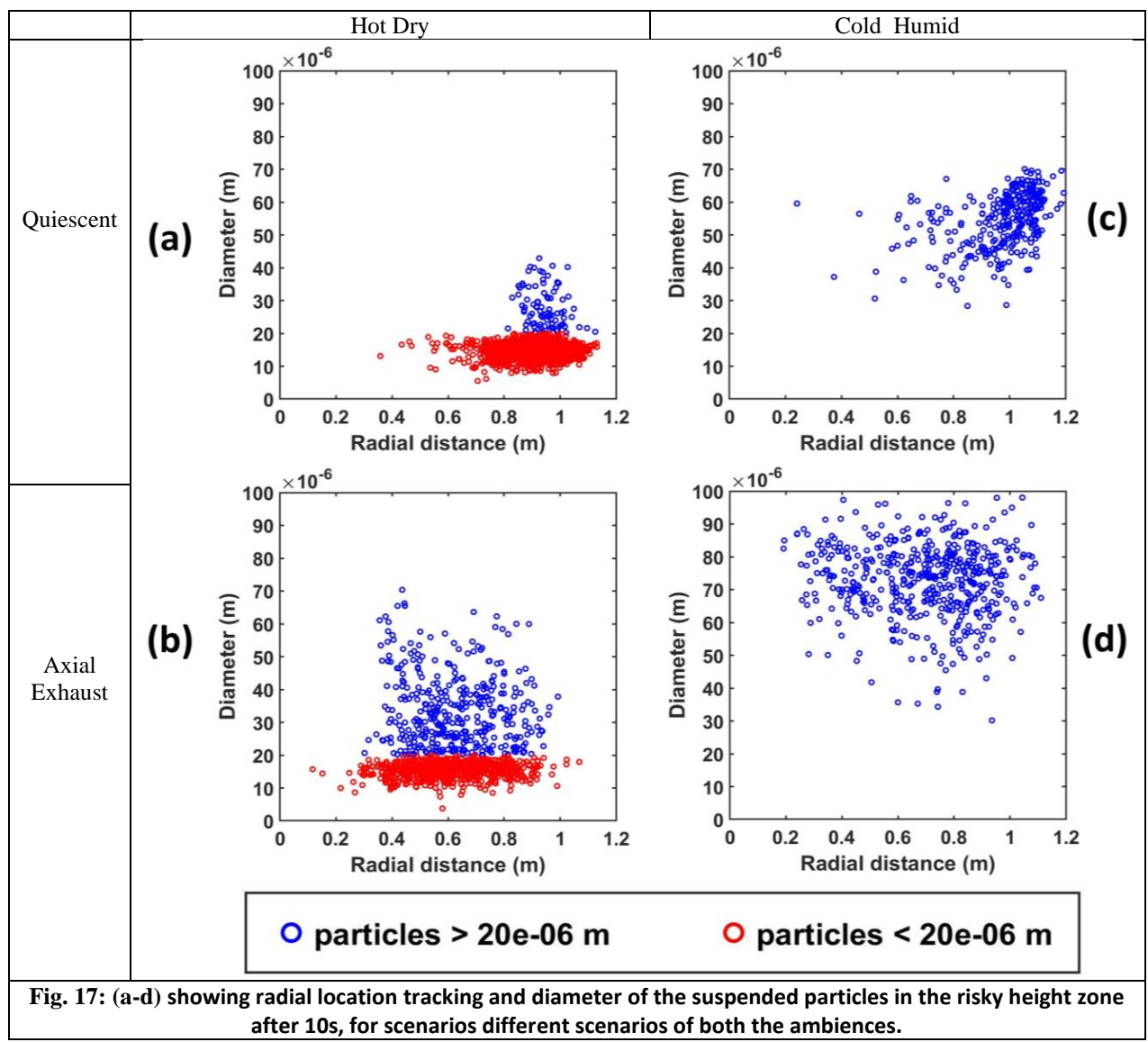

**Fig. 17:** (a-d) showing radial location tracking and diameter of the suspended particles in the risky height zone after 10s, for scenarios different scenarios of both the ambiences.

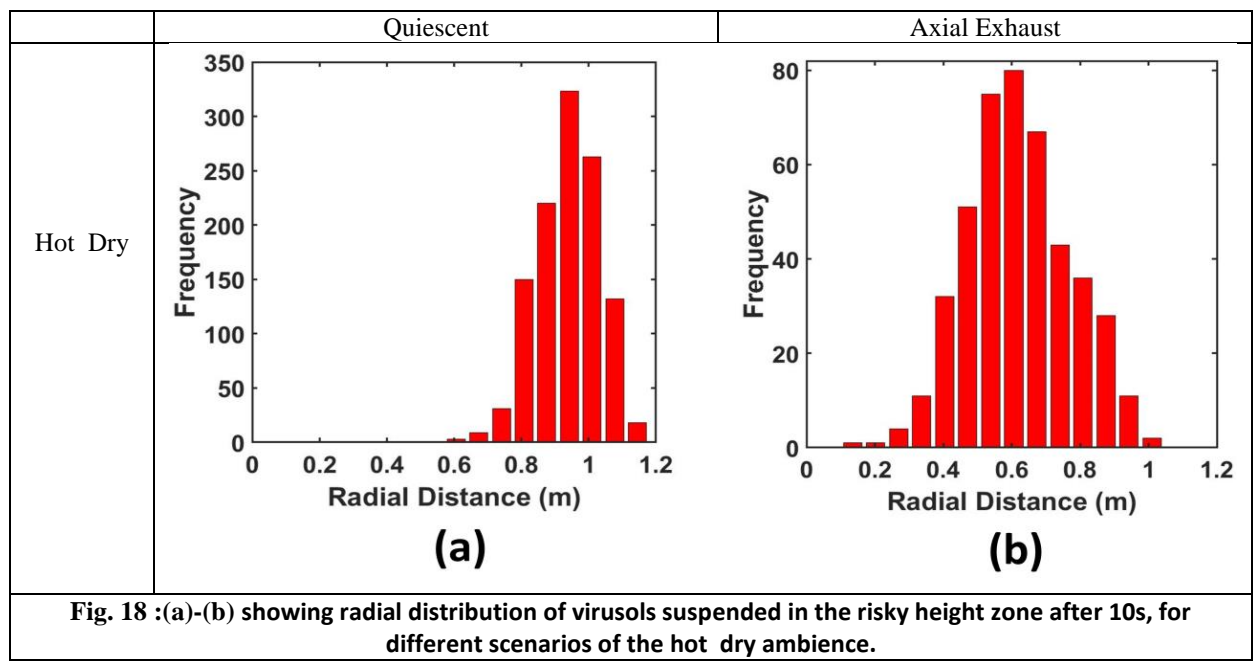

**Fig. 18 :(a)-(b) showing radial distribution of virusols suspended in the risky height zone after 10s, for different scenarios of the hot dry ambience.**

| Table 3: Table showing the most critical radial distance for various scenarios, for the hot dry ambience. | |
|---|---|
| **Scenarios** | **Distance (m)** |
| Scenario 1 | 0.95 |
| Scenario 2 | 0.6 |
| Scenario 3 | Entire Domain is Safe |

**V. CONCLUSION**

The transmission and evaporation of injected droplets in an elevator typically used in multi-storeyed residential or small enterprise buildings have been modelled. Three ventilation scenarios namely quiescent, axial exhaust and exhaust fan have been simulated within the elevator. As the droplets move according to the prevailing ventilation patterns, their evaporation is affected by prevailing air, velocity, temperature and humidity. In these investigations various scenarios have been explored to understand the droplet dynamics in the domain.

The simulation results show that, a quiescent condition in the elevator has a very high risk associated with it as significantly large percentage of droplets remain suspended in the domain in the risky height zone (0.8m to 1.8m) especially for the hot dry condition. For, the cold humid condition the droplets settle below the risky height zone at later stages due to their larger masses owing to negligible evaporation in such an ambience thus bringing down the risk significantly. Hence, the quiescent scenario turns out to be safer for the cold humid ambience (risk factor 7.22%, at 10s) as compared to the hot dry ambience (risk factor 40.32% at 10s).The introduction of force circulation ventilation in the form of axial exhaust jet or exhaust fan alleviates the situation by bringing down the risk factor as compared to quiescent scenario. This is attributed to the fact that with the introduction of forced circulation the percentage of droplets escaping out of the domain increases significantly. Although the axial exhaust jet fully alleviates the situation for hot dry condition (risk factor 24% at 10s) as compared to the quiescent scenario, the cold humid ambience produces different results. As the flow takes some time to develop in these cases for both the ambiences, the initially injected droplet streams goes downwards due to gravity forces which dominates over the upward pulling weak drag force but as the flow gets developed the droplets remain suspended in the risky height zone due to now present stronger drag force, whereas the droplets in quiescent scenario with cold humid ambience settle down below the risky height zone at latter stages of elevator travel bringing down the risk (risk factor 7.22%). Furthermore, it has been found that the risk in the exhaust jet scenario is less in the cold humid ambience than the hot dry ambience. The introduction of exhaust fan alleviates the situation for both the ambiences. The risk factor is significantly low for both the ambiences at all time instants and after 5.5s and 7.48s no droplets remain in the risky height zone for both the hot and cold ambiences respectively. Hence it is concluded that the exhaust fan is the best solution for mitigating. Also, it is found for all the ventilation scenarios the cold humid ambience is safer as compared to the hot dry ambience.

Although the forced circulation ventilation scenarios mitigate risk factor significantly than the quiescent scenario, in general, for forced circulation scenarios, the increased air velocity expedites the evaporation process of the droplets also the increase in turbulence of flow causing dispersion of droplets into areas having lesser mass fraction of water vapour also contributes to the increased evaporation rate especially for the hot dry ambience. The cold humid ambience produces negligible evaporation for all the ventilation scenarios. The increased evaporation rate increases the percentage of virusols among the suspended droplets. These virusols (diameter < 20µm), besides having very high viral loading also have the largest penetration in human trachea. The most critical radial distance (i.e. the radial distance having the highest concentration of virusols) for various scenarios is also obtained from our study. Other fellow passengers inside the elevator must always try to avoid this critical radial distance.

It is important to understand, that the scenarios that have been considered in this study are merely a fraction of a plethora of such likely real-life instances where the dimensions of the elevator, air supply and outflow slots, location of a fan or exhaust fan, velocity and speed, total time of elevator travel may be different from what are considered in this study. The passenger height and his/her positions within the elevator may also differ. The results are expected to be significant to these parametric variations and some of these situations may also present a substantial amount of threat in addition to the ones shown. Therefore, it is exhorted to take utmost precautionary measures while using an elevator.

**ACKNOWLEDGMENT**

We would like to thank the High performance computing cluster in the Technological Bhavan of Jadavpur University for allowing us to execute the simulations within a reasonable time period.

| NOMENCLATURE | |
|---|---|
| $C_p$ | Specific heat capacity of Eulerian phase (air) (Jkg$^{-1}$K$^{-1}$) |
| $C_{p,d}$ | Specific heat capacity of droplet (Jkg$^{-1}$K$^{-1}$) |
| $C_d$ | Coefficient of drag |
| $d_d$ | Diameter of droplet (m) |
| $D_{eff}$ | Effective diffusivity (m$^2$s$^{-1}$) |
| $D_{mol}$ | Molecular diffusivity in air (m$^2$s$^{-1}$) |
| $F_{\text{lift}}$ | Lift force on droplets(N) |
| $f_v$ | Mass fraction of water vapour in Eulerian phase |
| $H$ | Enthalpy (Jkg$^{-1}$) |
| $h$ | Convective heat transfer coefficient (Wm$^{-2}$K$^{-1}$) |
| $h_{fg}$ | Latent heat of vaporisation of droplet (Jkg$^{-1}$) |
| $k$ | Turbulent kinetic energy (Jkg$^{-1}$) |
| $k_t$ | Thermal conductivity (Wm$^{-1}$K$^{-1}$) |
| $k_{mt}$ | Mass transfer coefficient for droplet (ms$^{-1}$) |
| $m_d$ | Mass of droplet (kg) |
| $p$ | Static pressure of Eulerian phase (Nm$^{-2}$) |
| $P$ | Turbulent kinetic energy production (Nm$^{-2}$s$^{-1}$) |
| $RH$ | Relative humidity (percentage) |
| $T$ | Temperature (K) |
| $T_d$ | |

| | | |
|---|---|---|
| $\vec{u}$ | | Temperature of droplet (K) |
| | | Velocity of the Eulerian phase (ms$^{-1}$) |
| $\vec{u_d}$ | | Velocity of droplet (ms$^{-1}$) |
| ρ | | Density of Eulerian phase (kgm$^{-3}$) |
| $Y_d^s$ | | Mass fraction of nonvolatile component of droplet |
| $Y_d^l$ | | Mass fraction of volatile component of droplet |
| $ρ_d$ | | Density of droplet (kgm$^{-3}$) |
| $m_d^0$ | | Initial mass of droplet (kg) |
| $C_{p,s}$ | | Specific heat capacity of nonvolatile component of droplet (Jkg$^{-1}$K$^{-1}$) |
| $C_{p,l}$ | | Specific heat capacity of volatile component of droplet (Jkg$^{-1}$K$^{-1}$) |
| t | | Time (s) |
| μ | | Dynamic viscosity of Eulerian phase (kgm$^{-1}$s$^{-1}$) |
| μ$_t$ | | Turbulent viscosity of Eulerian phase (kgm$^{-1}$s$^{-1}$) |
| $Y_0^S$ | | Initial mass fraction of nonvolatile component of droplet |
| $m_{wl}$ | | Molecular wt. of volatile component in droplet(kg) |
| Sh | | Sherwood number |
| Nu | | Nusselt number |

## Appendix

**Grid Independence Study**

A 3D structured hexahedral-dominant mesh is used for the discretization of our computational domain. The mesh consists of approximately $5.8 \times 10^5$ cells. Mesh refinement is applied near the top mounting, outlets as well as other boundaries, but most importantly, sufficient refinement is applied near the passenger's mouth, from which the cough droplets are injected, to capture the droplet motion accurately. A smooth and gradual transition was also made from the very refined mesh near the mouth to the ambient mesh. A thorough grid independence study was conducted before selecting the above mesh. Three different mesh sizes; coarse (4,54,327 cells), medium (5,78,669 cells), and fine (6,95,027 cells) have been taken. The normalized eulerian field velocity $V_y/V_{inlet}$ in a hot dry ambience having the top mounting implemented as an axial inlet jet ventilation condition ($V_{inlet}$ = 0.56 m/s), is taken as the parameter, against which the results of the 3 different mesh are compared in Fig. 21 (b) . The normalized axial velocity ($V_y$) profile is taken along a line AB of 0.6 m length, the line AB being parallel to the X-axis. The line AB is situated on the plane CD. The plane CD is parallel to the elevator's floor and is at a height of 1 m above the elevator floor. The line AB and the plane CD and their positions inside the elevator domain are shown in Fig. 21 (a) . Besides the normalized axial velocity ($V_y$) profile, the velocity magnitude contours on plane CD are also compared in Fig. 19 (c) . The droplet dispersion and Risk Factors are compared in Fig. 20 (a) and (b) . From Figs. 19 (b), 19 (c) , 20(a), 20(b) , it can be seen that the results of the coarse mesh vary considerably from that of the medium and fine mesh and that the results of the medium and fine mesh have negligible difference. Since results become virtually grid independent with the medium and fine meshes, we adopted the medium mesh for our study. From this, it can be concluded that the medium mesh size considered here is adequate enough to capture the flow field correctly inside the elevator. Figures 21 (a) and 21 (b) show our finally adopted 3D mesh of 5,78,669 cells encapsulating the refinement near the boundaries and edges. Figure 21 (c) is showing a section of this mesh which portrays the gradual refinement near the mouth.

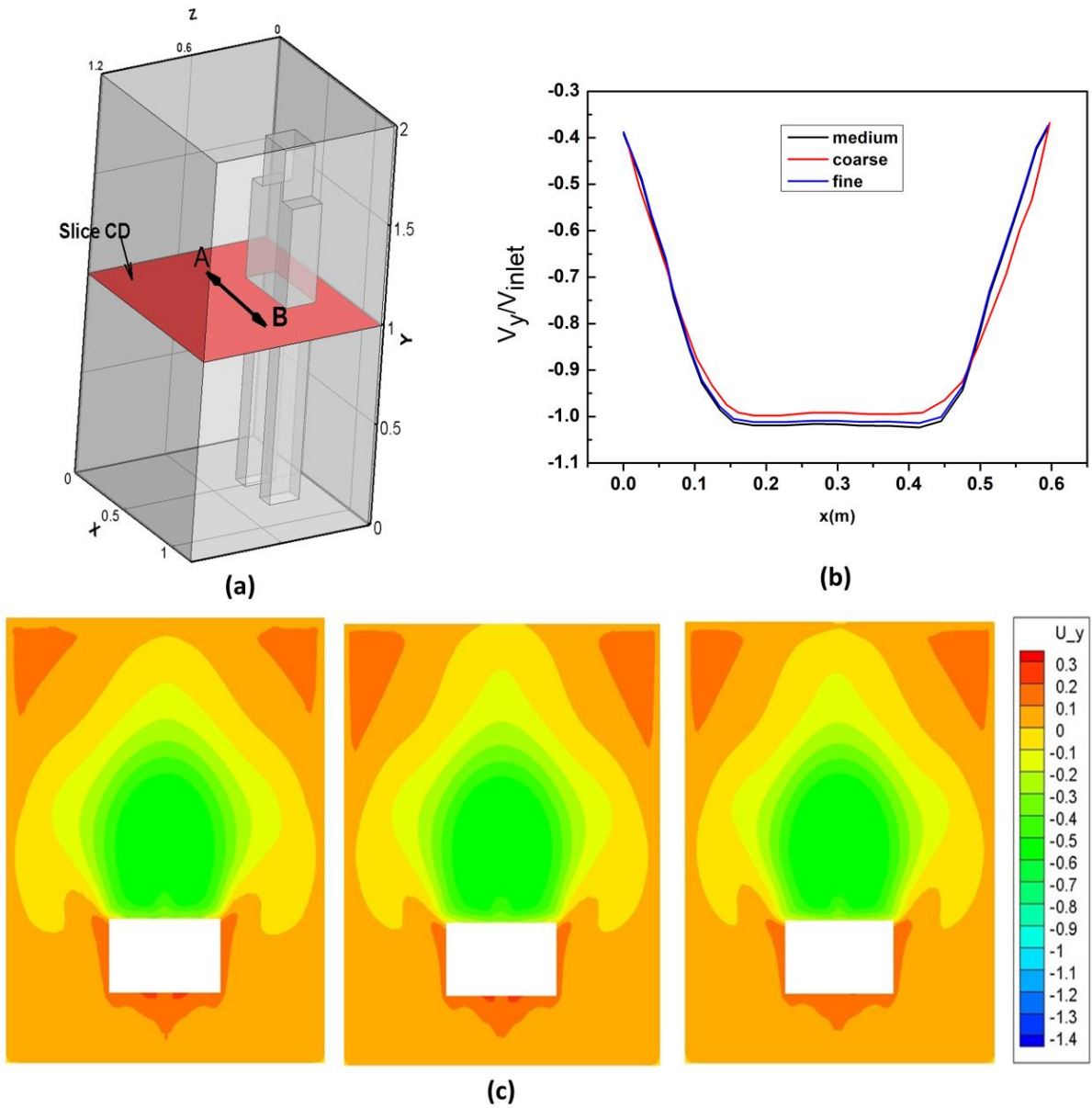

(a)

(b)

(c)

**Fig. 19 : (a) Location of slice CD and line AB**
**(b) Comparison of Velocity profile($V_y$) of different mesh sizes at line AB.**
**(c) Comparison of velocity contours of different mesh sizes at slice CD for coarse, medium and fine mesh sizes respectively from left to right**

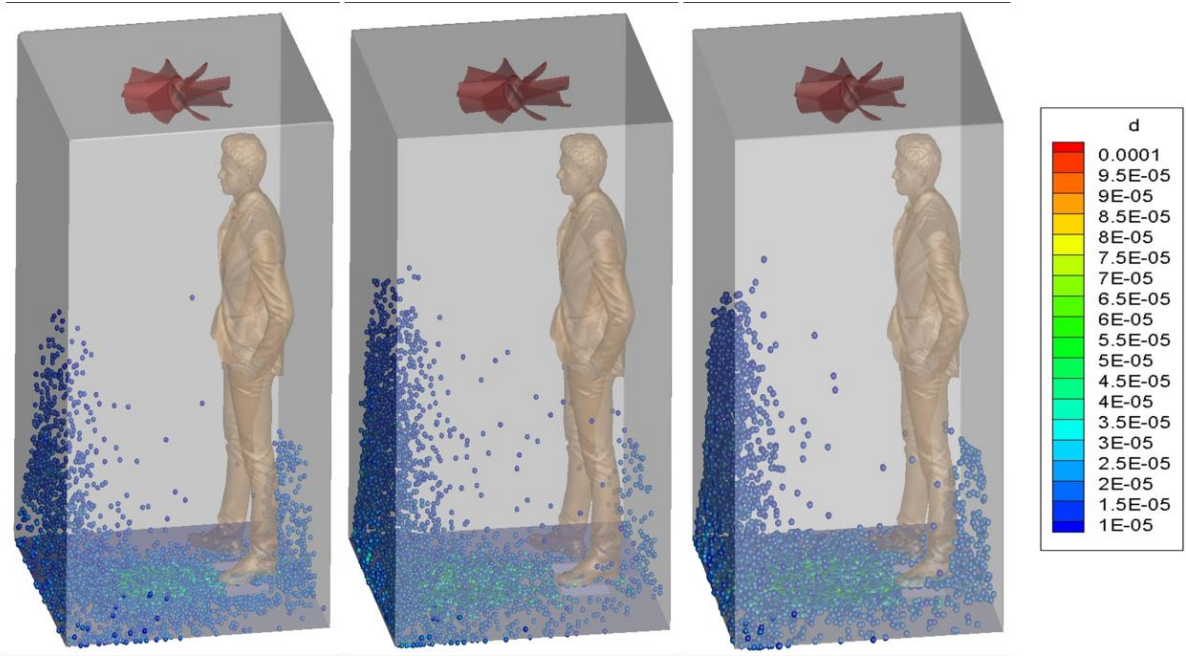

(a)

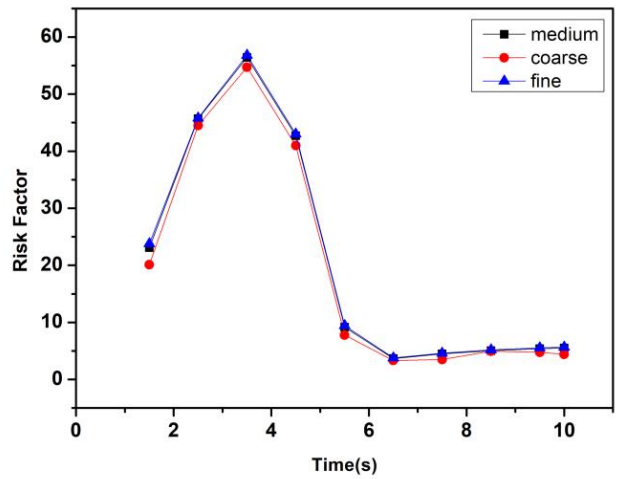

(b)

**Fig. 20: (a) Comparisons of droplet distributions for coarse, medium and fine mesh sizes respectively from left to right.**
**(b) Comparisons of risk factors for coarse, medium and fine mesh sizes respectively from left to right.**

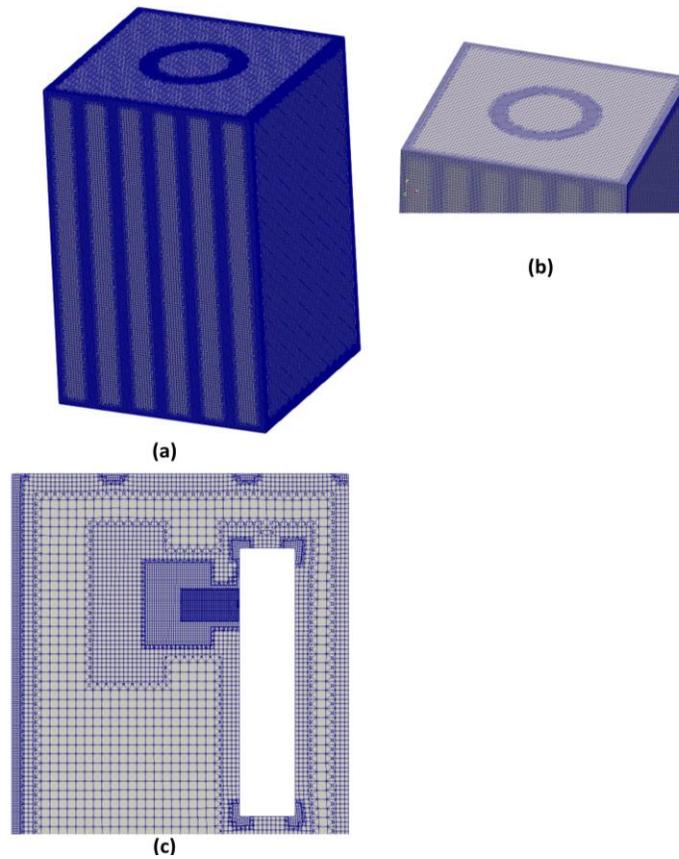

**Fig. 21 : (a) Adopted Mesh
(b) Refinements at various locations of the adopted mesh
(c) Mesh Section showing gradual refinement of mesh near mouth of the Passenger**